\documentstyle[12pt]{article}

\date  {{\small Revised June 5, 1998}}     

\def\be{\begin{equation}}
\def\ee{\end{equation}}
\def\bea{\begin{eqnarray}}
\def\eea{\end{eqnarray}}
\newcommand{\eq}[1]{eq.~(\ref{#1})} 

\def\S{{\mathcal S}}  

\def\too#1{\parbox[t]{.4in} {$\longrightarrow\\[-9pt]   {\scriptstyle #1}$}}

\def\blackbox{{\vrule height 1.3ex width 1.0ex depth -.2ex}\hskip 1.5
truecm}

\def\N {{I\!\! N}}

\newcounter{masectionnumber}
\newcommand{\masect}[1]{\setcounter{equation}{0}
  \refstepcounter{masectionnumber} \vspace{1truecm plus 1cm} \noindent
    {\large\bf \arabic{masectionnumber}. #1}\par \vspace{.2cm}
      \addcontentsline{toc}{section}{\arabic{masectionnumber}. #1}
    }
 \renewcommand{\theequation}
    {\mbox{\arabic{masectionnumber}.\arabic{equation}}}

    \newcounter{masubsectionnumber}[masectionnumber]
\newcommand{\masubsect}[1]{
    \refstepcounter{masubsectionnumber} \vspace{.5cm} \noindent
  {\large\em \arabic{masectionnumber}.\alph{masubsectionnumber} #1}
    \par\vspace*{.2truecm}

\addcontentsline{toc}{subsection}
 {\arabic{masectionnumber}.\alph{masubsectionnumber}\hspace{.1cm} #1}
    }
\newcommand{\startappendix}{ \setcounter{masectionnumber}{0} }
\newcommand{\maappendix}[1]{
    \setcounter{equation}{0}
  \refstepcounter{masectionnumber} \vspace{1truecm plus 1cm} \noindent
    {\large\bf \Alph{masectionnumber}. #1}\par \vspace{.2cm}

 \renewcommand{\theequation}
    {\mbox{\Alph{masectionnumber}.\arabic{equation}}}

  \addcontentsline{toc}{section}{\Alph{masectionnumber}. #1}
      }

\newtheorem{lem}{Lemma}[masectionnumber]
\newtheorem{thm}[lem]{Theorem}
\newtheorem{prop}[lem]{Proposition}

\newtheorem{df}[lem]{Definition}
\newtheorem{rem}[lem]{Remark}
\newtheorem{claim}[lem]{Claim}

\begin{document}
\title{\vspace*{-.35in}
On the Stability of the Quenched State
in Mean Field Spin Glass Models  }
\author{M. Aizenman ${}^{(a)}$
\qquad and \qquad P. Contucci ${}^{(b)}$\\ \hskip 1cm
\vspace*{-0.05truein} \\
\normalsize \it  ${}^{(a)}$ School of Mathematics,
Institute for Advanced Study, Princeton NJ 08540.
\thanks{Permanent address:  Departments of Physics and
Mathematics, Jadwin Hall,
Princeton University, P. O. Box 708, Princeton, NJ 08544.} \\
\normalsize \it ${}^{(b)}$ Department of Physics,
Princeton University,  Princeton NJ 08544.    }
\maketitle
\thispagestyle{empty}        
\newpage
\begin{abstract}
While the Gibbs states of spin glass models have been noted
to have an erratic dependence on temperature, one may expect
the mean over the disorder to produce a continuously varying
``quenched state''.
The assumption of such continuity in temperature
implies that in the infinite volume limit the state
is stable under a class of deformations of the Gibbs measure.
The condition is satisfied by the Parisi Ansatz, along with
an even broader stationarity property.  The stability conditions
have equivalent expressions as marginal additivity of the
quenched free energy. Implications of the continuity assumption
include constraints on the overlap distribution, which are 
expressed as the vanishing of the 
expectation value for an infinite collection of multi-overlap polynomials. 
The polynomials can be computed with the aid of a
{\it real}-replica calculation in which the number of replicas
is taken to zero.
\end{abstract}

\vfill
\newpage
\vspace{-1.2cm}
%
%
\masect{Introduction} \vspace{-.6cm}
\vskip .5truecm

We consider here the quenched state of the Sherrington-Kirkpatrick (SK)
spin glass model, and discuss some stationarity properties which seem to 
emerge in the infinite volume limit.

The SK spin glass model \cite{SK}  has spin variables $\sigma_i=\pm 1$,
$i = 1,\ldots, N$, interacting via the Hamiltonian
\be
H_N(\sigma,J) \ = \
- {1\over \sqrt{N}}
\sum_{1\le i<j\le N}J_{i,j}\sigma_i\sigma_j \ ,
\label{defenergy}
\ee
with $J_{i,j}$ independent normal Gaussian variables.

Sampling repeatedly spin configurations $\sigma^{(l)}$ from the space 
${\mathcal S}_N=\{-1,1\}^N$, distributed independently
relative to a common Gibbs state, one obtains a random matrix of
overlaps $q_{l,m} \ = \ q_{\sigma^{(l)},\sigma^{(m)} }$, 
with the overlap defined
for pairs of spin configurations $\sigma$ and $\sigma'$ as:
\be
q_{\sigma,\sigma'} \ = 
\ {1\over N}\sum_{i}\sigma_i\sigma'_i  \ .
\label{defoverl}
\ee
The quenched free energy, at a given temperature
$T = (k \beta)^{-1}$,  is  determined in this model by the
joint distribution of such overlaps for arbitrary number of
``replicas'' (copies of the spin system subject to the same
random interaction).

The purported solution of the model via the Parisi
ansatz \cite{MPV} has a number
of remarkable stability properties.   Our purpose is to discuss
some of these, starting from elementary continuity assumptions
without assuming the validity of the proposed solution.
Some results in a similar direction were  previously
obtained by F. Guerra~\cite{G}.

Following are the main observations presented here.
\begin{itemize}
\item[1)]
We identify a stability condition, in the sense of invariance of the
quenched state in the thermodynamic limit
under a broad class of deformations, which is satisfied
by the state corresponding to the Parisi solution.
\item[2)]
We show that a restricted version of the stability condition
is a consequence of a property
which make good physical sense, namely the continuity
of the quenched ensemble as function of the temperature.
\item[3)]
The restricted stability of the quenched state implies the vanishing of
expectation values for a family of overlap multireplica polynomials.
This accounts for some of the relations, though far from all, found
in the Parisi solution.
\item[4)]
We ask whether the stability condition singles the Parisi family of
states (the GREM models in the terminology of \cite{R}).
A considerably simplified version of such a question (concerning
the characterization of the the REM states) has a positive answer
\cite{AC2} (in preparation).
\end{itemize}

%
%
\masect{Invariance of the Parisi solution.}

Let us introduce the notation employed here for the quenched state,
and state a remarkable invariance property of the Parisi solution.

\masubsect{Notation for Quenched Ensembles}
In disordered spin systems we encounter two
distinct random structures: the spins, distributed  (in equilibrium)
according to
the Gibbs distribution, and the random couplings (and/or random
fields and other parameters) which affect the Gibbs state, making
it into a random measure.   We denote here by $< - >$, and in case
of possible confusion by  $< - >_J $, the expectation value over
the spins averaged with respect to the Gibbs state. An average over
the couplings is denoted by $Av( - )$.  The combined
{\em quenched average} is a double average,
denoted below by $E( - )$, over the spins and the disorder
(whose probability distribution is not affected
[in the quenched case] by the response of the spin system
to the random Hamiltonian).

Quantities of interest include:
\bea
E^{(N)}\left( q_{1,2}^2 \right)  \ & = &\
\frac{1}{N^2} \sum_{i,j} Av_N\left( \ <\sigma_i \sigma_j>^2_J \ \right)
\ = \ Av_N\left( \ <\sigma_1 \sigma_2>^2_J \ \right) + O(\frac{1}{N})
\; ,   \nonumber \\
\mbox{and} \; \; \;
& & \sqrt{N} Av_N\left(<\sigma_1 \sigma_2> \ J_{2,3} \ <\sigma_3 \sigma_1>
 \right)  \; .
\label{eq:example}
\eea
(where  the indices on 
$q_{1,2}$ have different meaning than those on  $\sigma_1 \sigma_2$).
The second example is seen among other terms in
${\partial \over \partial \beta}E^{(N)}\left( q_{1,2}^2 \right)$.
In expressions like this the factors $J_{i,j}$ can be integrated 
out, e.g., through the integration by part formula (for normal Gaussian 
variables) 
\be
Av \left( J f(J) \right) \ = \
Av \left( \frac{\partial}{\partial J} f(J) \right) \, . 
\label{eq:parts}
\ee
The expression can then be further reduced to an 
average of a suitable overlap monomial.  

{\em Overlap monomials} are functions of the form
$\prod_{1\le l,m\le K} q_{\sigma^{(l)},\sigma^{(m)} }^{n_{l,m}}$
defined over the product space $\S^{\otimes K}$, $\S$ being the 
spin configuration space.  Their expectation
values over the corresponding 
product measure on identical copies $\prod_{1 \le l \le K}< - >^{(l)}_{N,J}$
are denoted by the symbol
\be
 \ll \prod_{1\le i,j\le K}
  q_{\sigma^{(i)},\sigma^{(j)} }^{n_{i,j}} \gg_{N,J} \; , 
\label{rovlp}
\ee
and the full ``quenched average'' is denoted by
$E(-)$, e.g.:
\be
E^{(N)}(\prod_{1\le l,m\le K}q_{l,m}^{n_{l,m}})=
Av_N( \ll \prod_{1\le i,j\le K}
  q_{\sigma^{(l)},\sigma^{(m)} }^{n_{l,m}} \gg_{N,J} )  \; .
\label{qmsr}
\ee
We shall often omit the explicit reference to $J$, to 
other subscripts such as temperature, as well as to $N$. 
However, it should hopefully be clear from the context when do 
we refer to a finite system and when to the infinite volume limit.

Naturally, we are interested in the limit $N \to \infty$.   
There is no reason to expect (at low temperatures) 
convergence of the state $\ll - \gg_J$
at given realization of the random couplings $\{J_{i,j}\}$.
However it does not seem unreasonable to  expect convergence 
of the quenched averages of $\ll f(\sigma) \gg$, where $f(*)$ can 
be any  local function of the spins. 
We note that elementary compactness arguments imply that for any 
temperature there is a sequence $N_k$ which increases to $\infty$, 
for which the following limits exists simultaneously for all 
the overlap monomials 
\be
\lim_{k\to \infty} E^{(N_k)}(\prod_{1\le l,m\le K}
q_{\sigma^{(l)},\sigma^{(m)}}^{n_{l,m}})
=E(\prod_{1\le l,m\le K}q_{l,m}^{n_{l,m}}) \; ,
\label{limit}
\ee
Our discussion will concern relations among the monomial averages which 
would be valid in such limits, under a number of assumptions 
(which include the existence of the limit). 

\noindent {\bf Remark:} 
It is possible to develop a more complete setup for the formulation of 
the infinite volume limit of states of the SK model, in which the limit 
of the quenched averages is described in terms of a probability 
measure on  ${\mathcal M(S)}$ - the space  
of probability measures on  
${\mathcal S}=\{-1,+1\}^{\N}$ 
which is the space of configurations of an infinite spin system 
($\N$ being the set of natural numbers).
The elements of ${\mathcal M(S)}$ are random states incorporating  
the effects of quenched disorder. 
We shall, however, not pursue this line here.  

We shall invoke an additional element of structure: 
Gaussian fields $h(\sigma), K(\sigma)$ defined over $\S_N$ with the 
covariances
\bea
Av\left( h(\sigma^{(l)}) h(\sigma^{(m)}) \right) \ & = &  
\ q_{l,m}
\label{eq:h} \\
Av\left( K(\sigma^{(l)}) K(\sigma^{(m)}) \right) \ & = &  
\ q^2_{l,m} 
\label{eq:K} \\
Av\left( h(\sigma^{(l)}) K(\sigma^{(m)}) \right) \ & = &  \ 0  \; .
\label{eq:ind}
\eea
which are independent of
each other and of $J$, in the sense exemplified by the relation 
\be
E^{(N)}(e^{\sum_{j}\lambda_j h(\sigma^{(j)})}) \ = \ 
E^{(N)}( e^{ {1 \over 2} \sum_{n,m} \lambda_n \lambda_m q_{n,m} } ) \; .
\label{eq:indep}
\ee
(Analogous relation - with $q_{i,j}$ replaced by $q_{i,j}^2$, 
is assumed for $K$.)

In the SK model ($N< \infty$), 
a quantity like $h(\sigma)$ appears as the {\em cavity field}
associated with an increase in $N$, and $K(\sigma)$ 
can be found as representing the change in the action
corresponding to an increase in the temperature.
The two can be realized as:
\be
h(\sigma) \ = \ {1\over \sqrt{N}} \sum_{i=1}^{N} J'_{i} \sigma_i \; ,
\label{acca}
\ee
and
\be
K(\sigma) \ = \ {1\over N}\sum_{i<j} J''_{i,j} \sigma_i \sigma_j \; ,
\label{cappa}
\ee
where the $J'_i$ and $J''_{i,j}$ are normal Gaussian variables, 
independent form each other and from the couplings $J$ 
appearing in the Hamiltonian.  
Based on this example, we incorporate the averages over the fields 
$h, K$ under the symbol $Av(-)$, even when the average is at fixed 
$\sigma$.

\masubsect{Invariance of the Parisi solution}

The Parisi solution has the property that quenched averages
are not affected by the addition to the action of terms of
the form $F(K(\sigma), h(\sigma) )$, where $F(\cdot, \cdot)$
is any smooth bounded function.

To express the above stated property, let us consider the deformed
states
\bea
 < - >_{N,F(K,h)}\ & := & \ {< - \exp{\{F(K,h)\}}>_N \over
 <\exp{\{F(K,h)\})}>_N } \; , \nonumber  \\
 \mbox{  } \nonumber \\
  \ll - \gg_{N,F(K,h)} \ & = \ & \otimes_{l}< - >^{(l)}_{N,F(K,h)}
  \; ,
  \label{dinfpr}
\eea
and 
\be
E^{(N)}_{F(K,h)}( - ) \ = \ Av(\ll - \gg_{N,F(K,h)}) \; ,
\label{dmisura}
\ee
and let $E_{F(K,h)}(-)$ represent the corresponding limit, as
$N \to \infty$, for the expectation values of overlap monomials 
(assuming the limit exists).  

\begin{claim}
Assuming the validity of the Parisi solution, in the infinite volume 
limit at any temperature:
\be
 E_{F(K,h)}(  \prod_{1\le l,m \le K} q^{n_{l,m}}_{l,m}) \ = 
 E(\prod_{1\le l,m \le K} q^{n_{l,m}}_{l,m})   
\label{eq:parisi}
\ee
where the expectation value functionals are to be interpreted as the 
$N\to \infty$ limits of expectations of overlap monomials.
\end{claim}

We shall not verify this statement here -- the reader is invited
to do so from the solution which is discussed in~\cite{MPV},
references therein, and in \cite{R} -- instead we shall discuss 
the origin and 
consequences of a somewhat restricted invariance of this kind.

%
%
\masect{Continuity in the temperature and stability under deformation}

The broad stability of the quenched state expressed by \eq{eq:parisi}
has not yet been rigorously derived for the SK model.  We shall
now find that a somewhat restricted version of this condition follows
from a natural continuity assumption.

There is a significant difference between the spin--glass and
the ferromagnetic spin models in the effect of a change in
the temperature on the equilibrium state.
Reduction in the temperature amounts to increased coercion towards
the low energy states of the system.
If the ground state is unique, it is natural to expect the
equilibrium state to vary continuously  at low temperatures.
When there are only few ground states, one may expect
some discontinuities (as in the Pirogov-Sinai theory
\cite{PS}).  However, when there is a high multiplicity
of competing low energy states the result may be quite different.
Indeed it is reported 
that for a given realization
of the random Hamiltonian, the equilibrium state has a very erratic
dependence on the temperature. Nevertheless it may seem
reasonable to expect
that with the average over the  disorder, the quenched state
might vary continuously with $\beta$.

To illuminate the consequences of the continuity  assumption, let us
note that due to the addition law for independent Gaussian variables,
the Gibbs factor determining the equilibrium state at the
inverse--temperature $\beta + \Delta \beta$ can be presented as
a sum of two independent terms:
\be
(\beta + \Delta\beta) H(\sigma,J) \
\mathrel{\mathop{=}\limits^{\cal D}} \
\beta H(\sigma,J) + \delta(\beta) H(\sigma,\tilde{J}) \; ,
\label{eq:deltaH}
\ee
where $A \mathrel{\mathop{=}\limits^{\cal D}} B$ means that $A$ and
$B$  have equal probability distributions,
$\{ J, \tilde{J}\}$ are two independent sets of couplings, and
\be
\delta(\beta)=\sqrt{2{\beta\Delta\beta}+({\Delta\beta})^2}
\ee
With the action cast in the form \eq{eq:deltaH},
the modified state is seen to incorporate the
effects of a strong term (of the order of the volume)
pulling in some randomly chosen
directions, when the main term itself has many competing states.
The assumption of the continuity of the quenched state appears now
as less obvious, and it should therefore carry some
notable consequences.

A related observation can be made by considering the effects of
deformation of the state through the addition of a Gaussian field of
the type $K(\sigma)$ (\eq{eq:K}).  An easy computation
based on the fact that
\be
H(\sigma,\tilde{J}) \mathrel{\mathop{=}\limits^{\cal D}} \sqrt{N} K(\sigma)
\label{eq:HK}
\ee
shows that 
\be
\sqrt{\beta^2 + {\lambda^2\over N}} H(\sigma,J) \
\mathrel{\mathop{=}\limits^{\cal D}} \
\beta H(\sigma,J) + \lambda K(\sigma) \; ,
\label{eq:betalambda}
\ee
i.e. the deformation with a field $K$ is equivalent in a change of 
the order $O(\frac{1}{N})$ in the temperature:
\be
E_{\lambda K}^{(N, \beta)}(\prod q_{i,j}^{n_{i,j}}) \ = \ 
E^{(N, \sqrt{\beta^2 + {\lambda^2\over N}})}(\prod q_{i,j}^{n_{i,j}}) \; \; ,
\label{propri}
\ee
where the superscripts refer to the size
and the inverse--temperature, the subscript indicates a deformation
of the state in the sense of \eq{dinfpr} and, as for the \eq{eq:parisi}
the equality is understood when the two measures are restricted to the
quantities independent from the deformation variable $K$.

In the limit $N \to \infty$, the change in the temperature on the
right side vanishes.
This immediately leads to the following observation.
\begin{prop}
If in a certain temperature range
the quenched averages $E^{(N, \beta)}(\prod q_{i,j}^{n_{i,j}})$
are uniformly continuous in $\beta$,
as $N\to \infty$, and the
infinite--volume limit exist for the  quenched state, in the sense
of ~\eq{limit}, then the  limit $E^{(\beta)}(\prod q_{i,j}^{n_{i,j}})$ 
is stable under deformations by $e^{\lambda K}$, i.e.:
\be
E^{(\beta)}\left(\prod q_{i,j}^{n_{i,j}} \right) 
\ = \ E^{(\beta)}_{\lambda K}\left( \prod q_{i,j}^{n_{i,j}} \right).
\label{eq:stability}
\ee
for all the overlap monomials.
\label{thm:stability}
\end{prop}

Let us note that the assumptions made above imply also
another stationarity principle: the quenched state would
be invariant under the deformation induced by
$\ln 2\cosh\beta h$.  To see that, compare the state of $N+1$
particles with that of $N$.   The trace over the  ``last'' spin
yields  for
expectation values of functions of the ``first''  $N$ spins
\be
E^{(N+1, \beta)}\left(\prod q_{i,j}^{n_{i,j}}\right) \ =\
E^{(N, \tilde{\beta})}_{\ln 2\cosh\tilde{\beta}h}\left(\prod q_{i,j}^{n_{i,j}}
\right) \; ,
\label{propri2}
\ee
where $\tilde{\beta}=\sqrt{N\over N+1}\beta$ and $h(\sigma)$ is a
Gaussian field with covariance $q_{\sigma,\sigma '}$.
Under the stated assumptions, in the
thermodynamical limit the previous relation becomes:

\be
E^{(\beta)}\left(\prod q_{i,j}^{n_{i,j}}\right) \ =
\ E^{(\beta)}_{\ln 2\cosh\beta h}\left(\prod q_{i,j}^{n_{i,j}}\right) \; .
\label{propri3}
\ee

%
%
\masect{A logarithmic relation expressing the stability property}

The stability condition which follows from the above discussed
continuity assumption is indeed found among the properties of the
Parisi solution,
along with the more sweeping stability under deformations of the 
more general form,  as seen in \eq{eq:parisi}.   This invariance 
property can also be cast in the form of a ``logarithmic relation'',
which expresses an additivity property for the marginal
increments in the quenched free energy.

\begin{df}   We say that a random system, in the quenched state
$Av_N\left( \ll - \gg_N \right)$, has marginally-additive free energy
if for any finite collection
of independent Gaussian fields $K^{(1)}, K^{(2)},\ldots, K^{(l)}$
with the covariance \eq{eq:K}, and any smooth polynomially
bounded functions $F_1, F_2,\ldots, F_l$, the following limits
exist and satisfy
\be
\lim_{N \to \infty} Av_N \ln <\exp(\sum_{i=1}^{l} F_i(K^{(i)}))>_N =
\lim_{N \to \infty} \sum_{i=1}^{l} Av \ln <\exp F_i(K^{(i)})>_N \ ,
\label{eq:log}
\ee
where the indices label independent families of fields (not to be 
confused with replica indices).
\end{df}

Our main observation is that the above marginal additivity
of the quenched free energy
is equivalent to the {\em stability} of the quenched state
(in the infinite volume limit) expressed by
\be
E\left(\prod q_{i,j}^{n_{i,j}}\right) \ =E_{F(K)}\left(\prod q_{i,j}^{n_{i,j}}
\right) \ ,
\label{9}
\ee
where $F$ is an arbitrary smooth bounded function.

Let us note that
the expectation values of quantities involving any of the above,
can be evaluated by first integrating over the extraneous
Gaussian variables ($K$). Using Wick's formula, this
integration produces expressions involving the overlaps among
arbitrary number of replicas, as in:
\bea
E^{(N)}\left(  K_1\, K_2\, K_2'\, K_3'\right)\
&=& \ Av_N\left( <K>_N <KK'>_N <K'>_N \right)
\nonumber \\
& = &  \
E^{(N)}(q^2_{1,2}q^2_{2,3}) \; ,
\label{eq:dict}
\eea
where we defined $K_i=K(\sigma^{(i)})$.

Conversely, the averages of polynomials in replica overlaps,
as $q^2_{1,2}q^2_{2,3}$ in the above expression, can be easily
expressed through the expectation values of suitable products of
independent copies of the $K$ field. 

Clearly the stability implies the marginal additivity property
(of the free energy).
To prove the converse, we need to show that (assuming the two limits
exist)
\be
\lim_{N \to \infty}
Av_N \left[ <G(K)>_{N,F(K')} \right]^n \ = \
\lim_{N \to \infty}
Av_N \left[<G(K)>_N\right]^n \; ,
\label{eq:power}
\ee
for any integer $n$ and polynomial function $G$.
(The full statement takes a bit more general form
-- involving
products with different functions $G$ for the different copies
of the spin system, however by the polarization argument there
is no loss of generality in taking the same function $G$
for all the $n$ replicas.)
Let us note also that, by an elementary approximation argument,
it suffices to prove \eq{eq:power} for bounded functions $G$.

The logarithmic property (\ref{eq:log}) implies that for all
$\varepsilon$
\be
\varphi_N(\varepsilon) \equiv
Av_N \ln {{<\exp(\varepsilon G + F)>_N}
\over{<\exp(\varepsilon G)>_N<\exp(F)>_N}} \too{N\to \infty} 0 \;  .
\label{10}
\ee
For bounded $G$ the function $\varphi_N(\varepsilon)$ is analytical
in a strip containing the real axis uniformly in N, and the logarithmic
property \eq{eq:log} is equivalent to the vanishing of 
all the quantities $\varphi^{(n)}_N(\varepsilon)|_{\varepsilon=0}$ in the
infinite volume limit.
By an inductive argument, these conditions imply \eq{eq:power}:
first we observe that
\be
\varphi'_N(\varepsilon)|_{0}= Av_N(<G>_{N,F})-Av_N(<G>_N) \to 0,
\label{indu1}
\ee
which is the stability for the first power (one replica).
The second derivative gives
\be
\varphi''_N(\varepsilon)|_{0}= Av_N(<G>^2_{N,F})
-Av_N(<G^2>_{N,F})-Av_N(<G>_N^2)+Av_N(<G^2>_N) \to 0 \; .
\label{indu2}
\ee
Thus
\bea
\lim_{N \to \infty} Av_N(<G>^2_{N,F}) - Av_N(<G>_N^2) \ &=& \ 
\lim_{N \to \infty} Av_N(<G^2>_{N,F}) - Av_N(<G^2>_N)
\nonumber \\
\ & =&  \ 0 \; ,
\label{indu3}
\eea
where the last equality is by the first order equation, \eq{indu1}, applied
to the smooth function $G^2$.
Continuing in this fashion one may see that if stability is fulfilled up
to power $n$ it is fulfilled for power $n+1$.
\begin{rem}
The truncated expectations (cumulants) of order $p$ are generally defined by
\be
<-;p>^T=
{\partial^p \over \partial\lambda^p}\ln
<\exp (\lambda -)>|_{\lambda=0}  \;  .
\label{61}
\ee
In these terms the logarithmic relation is equivalent to:
\be
\lim_{N \to \infty}
Av_N <\sum_{i=1}^{l}F_i(K^{(i)});p>_N^{T} \ = \
\lim_{N \to \infty}
Av_N \sum_{i=1}^{l}<F_i(K^{(i)});p>_N^{T}
\label{eq:truncated}
\ee
for every integer $p$.  (The implication \eq{eq:log}
$\Rightarrow$ \eq{eq:truncated} is obvious.
In the other direction the proof can be based on the analyticity
argument indicated above.)
It might be noted that an equation like \eq{eq:truncated} cannot
possibly hold without the average {\em Av}, unless the Gibbs
state is
typically supported on a narrow collection of configurations
over which the overlap function
takes only the value  $q_{\sigma,\sigma'}=1$.
\end{rem}

Equations (\ref{eq:log}) and (\ref{eq:truncated}) have natural
counterparts for the more limited stability of the quenched state,
expressed by \eq{eq:stability}.  In that case
$F_i(K^{(i)})$ need by replaced by $\lambda_i K^{(i)}$, for
$i=2,3,\ldots$ (thought $F_1$ may still be left arbitrary.)

It is an interesting open question whether there are states
stable in the  limited sense which do not fulfill the stronger
stability condition.

%
%
\masect{Overlap polynomials with zero average}

We now turn to some of the implications of the stability condition
\eq{eq:stability} which was shown to follow from the continuity assumption.
Since the free energy and its derivatives are determined through the
distribution of the overlaps, it is natural to ask what consequences
does the stability property have in those terms.
As we shall see next, the implications include a family of relations
expressed as the vanishing of the expectation value of
an infinite collection of overlap polynomials.
From a combinatorial point of view, expressions with vanishing
expectation are constructed by applying a certain
operation to graphs representing overlap polynomials.

We shall use the notation encountered already in \eq{eq:dict},
where  $q_{1,2}$
indicates the overlap between two spin configurations sampled from two
different copies of the system, one in replica $1$ and the other in the
replica $2$, subject to the same random interaction.
Products of such
terms can be represented by labeled graphs, introduced below.
The expectation value does not depend on the particular labeling
of the different replicas, for instance
\be
E(q^2_{1,2}q^4_{2,3}q^2_{1,4})=E(q^2_{1,2}q^2_{2,3}q^4_{3,4}) \;  ,
\label{12}
\ee
where we omit, as in the rest of this section, 
the finite volume symbol $N$, unless otherwise specified.

Let now illustrate some consequences of the stability condition 
starting from the simple monomial
\be
Av(<K>^2_{\lambda K'}) \; .
\label{13}
\ee
Stability implies the vanishing of all the derivatives of this function 
of $\lambda$.  Let us proceede for a moment under the assumption, 
which is proven in the appendix,  that 
the limit of the derivatives in $\lambda$ equals the derivative of 
the limit, which under the stability condition is zero.  
The first derivative gives
\be
2Av_{N}€(<K>_{\lambda K'}(<KK'>_{\lambda K'}-<K>_{\lambda K'}<K'>_{\lambda K'}))
\label{a}
\ee
which, at $\lambda=0$, vanishes for the trivial reason of parity, at 
$\lambda=0$.
On the other hand, the second derivative yields
\begin{eqnarray}
2Av_N(<KK'>^2_{\lambda K'}-4<K>_{\lambda K'}<K'>_{\lambda K'}<KK'>_{\lambda K'}
+3<K>^2_{\lambda K'}<K'>^2_{\lambda K'}\nonumber \\
+<K>_{\lambda K'}<KK'^2>_{\lambda K'}
-<K>^2_{\lambda K'}<K'^2>_{\lambda K'}).
\label{14}
\end{eqnarray}
At $\lambda = 0$ the last two terms cancel and the expression
reduces to:
\be
2 E^{(N)} (q_{1,2}^4-8q^2_{1,2}q^2_{2,3} + 6q^2_{1,2}q^2_{3,4} ) \;  .
\label{15}
\ee

If the quenched state is stable in the sense of \eq{propri} 
the above expression tends to zero in the 
thermodynamic limit.

As was mentioned already, the stability property is satisfied by the Parisi
solution, and hence this relation, as well as those of higher order
derived below,  are satisfied there.
The particular case of vanishing of \eq{15}  was
recently derived (for almost every $beta$) without any assumptions
in ref.~(\cite{G}).
One may also note that the vanishing of \eq{15} is also the lowest non-trivial
identity of those listed in \eq{eq:truncated} (corresponding to
$p=4$, $F(K)\equiv K$).

Let us present now a systematic approach for the derivation of 
other such relations.

One may use a graphical representation in which a monomial of the
form $q^2_{1,2}q_{2,3}$ is identified with a graph whose vertices
are the replica indices $\{1, 2, 3\}$ and the edges correspond to the
overlaps, $q_{i,j}$.
Such a graph will be indicated by the symbol
$(1,2)^2(2,3)$.    Furthermore, we shall consider also products
involving an additional Gaussian field ($K$).  The graphical
representation of that factor is a half-edge, represented
by a singleton.  I.e.,  $(1,2)(2)$ and $(1,2)(3)$ correspond to
$q_{1,2}K_{2}$ and $q_{1,2}K_{3}$.

We shall use a product ``$\cdot$'' which acts in the space of graphs
as {\em composition} combined with {\em contraction}, where possible, of
the two unpaired legs.   The notion may be clarified by the following
examples:
\bea
(1,2)\cdot(1,2)(2) & =& (1,2)^2(2)  \nonumber  \\
(1,2)(3)\cdot(4) & = &  (1,2)(3,4)    \\
\eea
Terms of the form $(1,1)$ can be omitted, since in our case
$q_{1,1}=1$.

The above product turns out to be commutative but not
associative. The {\it order} of a graph is defined as
$2\times\mbox{\em number of edges}$, with half-edges counting  as $1/2$.
Let  $W_k$ denote the space of formal linear combination of graphs
of order $k$.  For $G\in W_k$ we denote by $Q_G$ the corresponding
element of the overlap algebra.

We define $\delta: W_k \rightarrow  W_{k+1}$ as the linear operator
which acts on single graphs by
\be
\delta G=\sum_{v\in V(G)}\delta_v G,
\label{der}
\ee
where $V(G)$ is the set of vertices of $G$, and
\be
\delta_v G= G\cdot (v) -G\cdot (\tilde{v}),
\label{deri}
\ee
where $\tilde{v}$ is a new vertex not belonging to $G$.
E.g.,  $\delta_1 (1,2)= (1,2)(1)-(1,2)(3)$,
$\delta_1 (1,2)(1)=(1,2)(1)\cdot(1)-(1,2)(1)\cdot(3)=(1,2)-(1,2)(1,3)$.

Following is the pertinent observation.
\begin{prop}
For any measure of the type $E^{(N)}(-)=Av_N(\ll - \gg_N)$ and a deformation
defined in (\ref{eq:stability})
\be
{\partial^2\over \partial \lambda^2} E^{(N)}_{\lambda K}(Q_G)|_{\lambda=0}=
E^{(N)}(Q_{\delta^2 G}) \; ,
\label{prop:16}
\ee
for all the elements $Q_G$ of the overlap algebra and every $N$.
\end{prop}

The proof of Proposition 5.1 is straightforward.
The operation $\delta$  is the graphical counterpart of the usual derivative
with respect to the parameter $\lambda$ in the Boltzmann weight
(where it appears in $\lambda K$).
Such a derivative produces a {\it truncated correlation} expressed  in
the rule (\ref{deri}). The (\ref{der}) is nothing
but the Leibnitz rule for derivative of products. The first
differentiation produces a sum of monomials, each containing an unpaired
centered Gaussian variable ($K$) of zero mean.
The second derivative produces another unpaired variable,
which is contracted
with the previous one via the Wick rule. (This contraction motivates the
product introduced above.) 

\begin{prop}
If in a certain temperature range
the quenched averages $E^{(N, \beta)}(Q)$
are uniformly continuous in $\beta$,
as $N\to \infty$, and the
infinite--volume limit exist for the  quenched state, in the sense
of ~\eq{limit}, then 
\be
E(Q_{\delta^2 G}) = 0
\ee
for every element of the overlap algebra.
\end{prop}

The uniform continuity in $\beta$ implies the stability for 
deformation \eq{eq:stability} which means, in particular, that
the limit is a constant in $\lambda$. To prove the theorem we have to show
that we can interchange the thermodynamical limit with the repeated 
differentiation w.r.t. $\lambda$, in \eq{prop:16}. 
This is shown to be true in the appendix, 
through uniform (in N) bounds on the k-th derivatives of expectations 
of overlap monomials.

Following is a related statement which yields a somewhat stronger 
conclusion (suggesting a numerical test), which is
derived under a stronger assumption.  

\begin{prop} In the SK model, at finite values $N$,
\be
{\partial\over \beta\partial\beta}E^{(N)}(Q_G)=N
{\partial^2\over \partial \lambda^2} E^{(N)}_{\lambda K}(Q_G)|_{\lambda=0}=
N E^{(N)}(Q_{\delta^2 G}),
\label{17}
\ee
for every element of the overlap algebra and every $N$. In particular
if in a certain range of $\beta$ the quantities
${\partial\over \beta\partial\beta}E^{(N)}(Q_G)$ are uniformly bounded
in $N$, and the thermodynamic limit exist in the sense of \eq{limit},
then for all the elements $Q_G$ of the overlap algebra:
\be
E(Q_{\delta^2 G})=O(1/N) \; .
\label{170}
\ee
\end{prop}

The proof of the first equality can be obtained computing the second 
derivative with respect to $\lambda$ of equation \eq{propri} 
at $\lambda=0$. The second equality is \eq{prop:16}.

Analogous statements hold for other mean field spin glass
models, with the $p$-spin interaction Hamiltonian (\cite{D})
\be
H(\sigma,J)=
-\sum_{1\le i_1<\cdots< i_p\le N}J_{i_1,\cdots,i_p}\sigma_{i_1}\cdots
\sigma_{i_p},
\label{genep}
\ee
where $J_{i_1,\cdots,i_p}$ are Gaussian variables, rescaled
so that the Hamiltonian covariance is
\be
Av H(\sigma,J)H(\sigma',J)=N q^p_{\sigma,\sigma'}.
\label{pcova}
\ee
(Under the above calling, $H$ reaches values of order $N$.
The corresponding choice for the  field $K$ is Gaussian with the
covariance
$Av(K(\sigma)K(\sigma '))=q^p_{\sigma, \sigma '}$.
The definition of $\delta$ is unchanged.

%
%
\masect{Computation with real replicas}

In this section we give several characterizations of the
overlap polynomials of the form $Q_{\delta^2 G}$ for which we prove
that under certain conditions they have zero mean. 

The main result is a formula
which permits to compute the polynomials from a quadratic expression
in the number $r$ of {\it real}-replicas, evaluated at $r=0$.
To state it, let  $M_r=\sum_{i\ne j=1}^{r}q^2_{i,j}$, for all integers
$r\ge 1$.  Let $E(-)$ be an  expectation value functional,
on the algebra of overlaps, which depends only on the
graph structure of the overlap monomials (i.e., is independent of
the choice of labels).  Then the quantity $E(Q_G M_r)$ is quadratic
in $r$.   In the following proposition, we refer to the {\em
polynomial extension} of this function to all real $r$.

\begin{prop}  For any expectation value functional $E(-)$, as above,
\be
E(Q_{\delta^2 G})=E(Q_G M_r)|_{r=0}.
\label{eq:replica}
\ee
where the quantity $E(Q_G M_r)$ is first computed
for $r$ large enough so that all the indices appearing in
$Q_G$ do appear also in $M_r$ and $r>|G|+1$.
\label{prop:replica}
\end{prop}

To illustrate the statement, let us take: $G=q^2_{1,2}$.
In this case, the left side of \eq{eq:replica}
is given by \eq{15}  and the right side is:
\be
E(q^2_{1,2}M_r)=2E(q^4_{1,2})+4(r-2)E(q^2_{1,2}q^2_{2,3})
+(r-2)(r-3)E(q^2_{1,2}q^2_{3,4})  \;  .
\label{exem}
\ee
The two coincide at $r=0$ (defined by polynomial extension).

The proof proceeds through the explicit computation of
the left and right sides of \eq{eq:replica}.

\begin{lem}
If the number of vertices in $G$ is $l$ then
\be
\delta^2 G = \sum_{v\ne v'}G\cdot (v,v')-2l\sum_{v}G\cdot (v,\tilde{v})+
l(l+1)G\cdot (\tilde{v},\tilde{v}'),
\label{espgr}
\ee
where $v$ and $v'$ are summed over the set of vertices of
$G$ and $\tilde{v}$ and $\tilde{v}'$ denote a pair of added vertices.
\end{lem}
This is a rather explicit expression for the polynomials corresponding
to  a given graph $G$. Two examples are:
\bea
\delta^2 (1,2)(3,4) & = &
4(1,2)^{2}(3,4)\ +\ 8(1,2)(2,3)(3,4) -
\nonumber \\
& & -32(1,2)(2,3)(4,5) \ + \ 20(1,2)(3,4)(5,6) \; ,
\label{twopi}
\eea
and
\bea
\delta^2 (1,2)(2,3) & = &
4(1,2)^{2}(2,3)+2(1,2)(2,3)(3,1)\ -\ 12(1,2)(2,3)(3,4) +
\nonumber \\
& & + 12(1,2)(2,3)(4,5)\ -\ 6(1,2)(2,3)(2,4) \; .
\label{thevi}
\eea

\noindent{\bf Remark} In the above example $\delta^2 G$
is a polynomial expression with integer coefficients
whose sum is zero.  That property is shared by  $\delta^2 G$
of arbitrary monomials $G$.

To prove the lemma we note that by the definition of $\delta$
\be
\delta G= \sum_{v}G\cdot (v)- lG\cdot (\tilde{v}) \; .
\label{lem11}
\ee
Applying this rule twice
\bea
\delta^{2}G\ & =\   & \sum_{v,v'}G\cdot (v,v')-l\sum_{v}G\cdot (v,\tilde{v})
-l\sum_{v'}G\cdot (v',\tilde{v})-l\sum_{v'}G\cdot (\tilde{v},\tilde{v})\\
\nonumber
& & +  l(l+1)G\cdot (\tilde{v},\tilde{v}'),
\label{lem12}
\eea
which coincides with \eq{espgr} since  $(v,v)=1$ for every $v$.

\begin{lem}
If all the replica indices appearing in $Q_G$ are contained in the entries
of the matrix $M_r$ and $r>l+1$ then
\bea
E(Q_G M_r)\ =\ \sum_{
\begin{array}{c}
{\footnotesize 	v, v' \in V(G) } \\
{\footnotesize	v\ne v' }
\end{array} }
E(Q_{G\cdot (v,v')})
\ + \ 2(r-l)\sum_{v\in V(G)} E(Q_{G\cdot (v,\tilde{v})}) \ \nonumber
\\
+\ (r-l)(r-l-1)E(Q_{G\cdot (\tilde{v},\tilde{v}')}).
\label{rrep}
\eea
\end{lem}

This formula is an elementary consequence of the fact that the measure
$E(-)$ depends only on the isomorphism type of the graph associated to a
given overlap monomial.  The first sum on the right side of  \eq{rrep}
corresponds to  those overlap terms in
$M_r$ which involve only the replicas appearing in $G$, the other two
sums are split according to whether the number of vertices not in $G$
is $1$ or $2$.

The two previous lemmas prove Proposition~\ref{prop:replica}.
\hfill\blackbox

\masect{Comments} \vspace{-.6cm}
\vskip .5truecm

We have seen that elementary continuity assumptions
on the quenched state, imply a stability property for the
infinite volume limit of the SK (and other mean-field) 
spin-glass models.
A particular implication is the 
vanishing of the expectation values of certain 
multi-overlap polynomials, which form an infinite dimensional family.
We also saw a related condition expressed through an explicit decay rate
for the expectation values of suitable quantities.

These observations are consistent with the Parisi theory.
However, the family of identities
discussed here does not yet permit the reconstruction of the joint
probability distribution from that of a single overlap, as is the
case under the Parisi Ansatz.

It has been pointed out that within the replica-symmetry-
breaking approach the vanishing of the expectation
values of the polynomials discussed here ($\delta^2 G$) requires
only the so called ``replica--equivalence'' assumption, which says that
in the matrix $Q$ (defined in \cite{MPV}) each row is a
permutation of any other.  We thank I. Kondor and M. Mezard for 
calling our attention to this point.  
See \cite{P} for a recent account on
replica-equivalence.

An interesting question is whether the stability property is the
stationarity condition for some variational principle.
This is related to the main question which emerges
at this point, which is whether  stability
implies the GREM state structure \cite{R}.
We study a restricted version of this question
in a separate paper \cite{AC2}.

\startappendix
\maappendix{Appendix}

In this appendix we show that the limit $N\to \infty$ can be 
interchanged 
with the differentiation, in formula (\ref{prop:16}). 
This is seen in two steps.  The first is a general criterion.
\begin{thm}
Let  $F_{n}(\lambda)$  be a sequence of functions 
defined  over the interval $ \lambda \in [-1,1] $, which:

\noindent
a)  converge pointwise:
\be
F_{n}(\lambda) \too{n \to \infty} G(\lambda) \qquad 
 \mbox{ for all $\lambda \in [-1,1]$,}
\ee
b) have uniformly bounded  derivatives up to order $m+1$, i.e. satisfy 
\be
|{d^{k} \over d\lambda ^{k} } F_{n}(\lambda) | \ \le \ Const.  
\ee
for all $\lambda \in [-1,1]$ and $k=1,\ldots, m+1$.   

Then for $ k=1,\ldots,m$ the derivatives also converge (pointwise and 
uniformly), the limit is differentiable, and     
\be
{d^{k} \over d\lambda ^{k} } F_{n}(\lambda) \too{n \to \infty} 
         {d^{k} \over dx^{k} } G(\lambda)
\ee
\end{thm}
We omit the proof of this basic criterion.  (For $m=1$ it can be proven 
by using uniform approximations for ${d \over d \lambda}F_n(\lambda)$ in 
terms of the differences $[F_n(\lambda + \epsilon) - 
F_n(\lambda)]/\epsilon$, and the rest is by induction.)

The stability stated in Proposition 3.1 amounts to an (a) 
type condition for $F_{n}(\lambda) = E^{(N)}_{\lambda K}(Q)$ where $Q$ 
is any overlap monomial.  The limit is a constant function.  
The above principle will allow us to conclude that 
under the assumption of Proposition~(3.1) $E(Q_{\delta^{2} G}) = 0$, 
as soon as we show that the derivatives are uniformly bounded 
(i.e., establish condition (b)).  Following is a detailed version of 
that statement.   

\begin{prop}
Let $Q$ be an overlap monomial which involves $r$ replicas. 
Then, for any $N < \infty$,  
\be
 |{d^{k} \over d\lambda ^{k} }
  E^{(N)}_{\lambda K}(Q) | \ \le \ 2 r^{k+1} \ k! \ (k+1)^{k+1} 
  e^{(1+\lambda)^2 + \lambda^2}  \; .
\ee
\end{prop}

\noindent {\bf Proof:} 
The expectation value of $Q$ are expressed as 
$E^{(N)}_{\lambda K}(Q) \ = \ Av_N \ll Q \gg_{N,\lambda K}$ 
with a finite product 
$\ll - \gg_{\lambda K} =  
   \otimes_{i=1,\ldots,r}< - >^{(i)}_{\lambda K} \; $.
Using standard formula for the derivative, standard 
bounds on the truncated correlations (see for instance Theorem II.12.6 in 
\cite{S}), and the fact that $|Q|\le 1$

\begin{eqnarray}
	|{d^{k} \over d\lambda ^{k} }E^{(N)}_{\lambda K}(Q) | \ 
  & = & \ | Av_N\left( \ll 
  Q; \sum_{i=1}^{r} K(\sigma_i);
  \cdots; \sum_{i=1}^{r} K(\sigma_i) \gg_{N,\lambda K} \right) |
	\nonumber   \\
  & \le & \  (k+1)^{k+1} 
\ Av_N \left( \ll \left| \sum_{i=1}^{r} K_i \right|^{k} 
\gg_{N,\lambda K} \right)
	\nonumber  \\ 
  & \le & \ (k+1)^{k+1} \ r^{k+1} 
\ \  Av_N\left( <|K|^k>_{N,\lambda K} \right)   
  \nonumber 
	\end{eqnarray}  
Since $K$ can be arbitrarily large, we 
face here a minor version of the ``large field problem''.  However, 
it can be resolved by the following estimate, 
 \begin{eqnarray} 
 {1 \over k!} Av_N\left( <|K|^k>_{N,\lambda K} \right)
&\le &  \nonumber \hfill \\ 
  & \le &  \     \ 2 Av_N\left( <e^{K}>_{N,\lambda K} \right)   \  = 
  \ 2 Av_N\left( { <e^{(\lambda +1) K}>_N
     \over  <e^{\lambda K}>_N} \right)  \nonumber \\
  & \le  & \ 2 \left( Av_N \left[<e^{(\lambda +1) K}>^2_N \right]^2 
 \right)^{1/2} \times 
\left( Av_N  \left[ <e^{\lambda K}>_N \right]^{-2} \right)^{1/2}  
\nonumber 
\end{eqnarray}  
For the first factor, an elementary calculation gives the 
bound $e^{(1+\lambda)^2 }$. For the second factor we get 
\be
Av_N  \left[<e^{\lambda K}>_N \right]^{-2} \le   
Av_N\left( \exp\left[ - 2 \lambda <K>_N \right] \right)
 \ \le \ e^{2 \lambda^2} 
\ee
where we used first the Jensen inequality and then the fact that 
$<K>$ is a gaussian variable of covariance  $\le 1$.\hfill\blackbox     



\noindent {\large \bf Acknowledgments\/}
This work was supported by the NSF Grant PHY-9512729 and,
at the Institute for Advanced Study, by a grant from
the  NEC Research Institute.


\end{document}